# CLOUD COMPUTING AVOIDS DOWNFALL OF APPLICATION SERVICE PROVIDERS

Kathleen Jungck[1] and Syed (Shawon) M. Rahman, PhD[2].

[1]Information Assurance and Security, Capella University, Minneapolis, USA
KJungck@CapellaUniversity.edu
[2]Assistant Professor, University of Hawaii-Hilo, Hilo, USA
and Adjunct Faculty, Capella University, Minneapolis, USA
SRahman@Hawaii.edu

### ABSTRACT

*Businesses have become dependent on ever increasing amounts of electronic information and rapid transaction speeds. Experts such as Diffie [1] speculate that the end of isolated computing is at hand, and that within the next decade most businesses will have made the shift to utility computing. In order to cut costs while still implementing increasingly complex Information Technology services, many companies turned to Application Service Providers (ASPs). Due to poor business models, over competition, and poor internet availability and bandwidth, many ASPs failed with the dot com crash. Other ASPs, however, who embraced web services architecture and true internet delivery were well placed as early cloud adopters. With the expanded penetration and bandwidth of internet services today, better business plans, and a wide divergence of offering, cloud computing is avoiding the ASP downfall, and is positioned to emerge as an enduring paradigm in computing.*

## 1. INTRODUCTION

Information technology (IT) is no longer merely a strategic advantage for businesses, but rather something the enterprise is now dependent upon. IT is changing and evolving at such a rapid pace than many businesses, especially small to medium enterprises (SMEs), have a difficult time keeping pace with costly, increasingly complex computing technology. Partial outsourcing, pioneered by Application Service Providers (ASPs), floundered due to a convergence of problems including poor business models, under capitalization, poor architecture implementation, lack of divergence in offerings, poor internet access and bandwidth, and too much competition following the 2001 dot com bubble. A few successful ASPs redesigned their offerings and survived to enter first the Software as a Service (SaaS) market, and later emerged as early competitors in cloud computing.

Unlike its predecessor, cloud computing offers multiple models. Instead of simply delivering applications in a client-server environment like early ASP models, cloud computing offers a true web services delivery. Cloud computing also offers scalable hosting infrastructure services and platform development services as well as utility model pricing. With the wide variety of services available, cloud computing overcomes many of the early ASP model's greatest drawbacks and while offering attractive Return on Investment (ROI) opportunities. Federal government adoption by the General Services Administration (GSA) as well as promotion of cloud computing with select, certified vendors, will prompt greater adoption of cloud computing [2].

## 2. BACKGROUND STUDY

Information has become the life blood of our society, with companies dependent on ever increasing amounts of electronic information and rapid transaction speeds in order to carry out their business [3]. The computing paradigms that provide and process this flow of information have changed multiple times since the inception of the first computer and continue to evolve at an ever increasing pace [1],[4],[5]. Yet another new paradigm, cloud computing, is moving to the forefront [5],[6],[7].

Increasing costs, fast paced technological changes, and a lack of highly trained Information Technology (IT) staff have made it difficult for small to medium enterprises (SMEs) to compete with the increasingly complex computing technology employed by larger enterprises. In 1989, Kodak outsourced their IT services to a third party, but soon discovered that complete IT outsourcing exposed vulnerabilities to proprietary data and processes [8]. Partial outsourcing, in the form of Application Service Providers (ASPs) began to pick up momentum [8],[9]. After a rocky start in the wake of the 2001 dot-com bubble, ASPs redesigned their offerings and established a viable foothold among SMEs [10].

Cloud computing brings another form of outsourcing to the table, in essence providing "supercomputing to the masses" [11]. Some ASPs are positioned to take advantage of the new cloud computing paradigm, which some experts consider to be an extension of the ASP model [7]. Other ASPs may have to undergo radical shifts within their business models in order to move into the cloud.

### 2.1. What is an ASP?

The term Application Service Provider, or ASP, was first used in 1996 when Jostein Eikeland recommended to a Norwegian bank that they run desktop applications remotely [12]. Eikeland went on to form Telecomputing, Inc., one of the largest companies in the ASP market [12].

An ASP assumes the responsibility of purchasing, hosting, and maintaining a software application on either its own infrastructure, one leased from a third party, or on the client's own infrastructure. The ASP then provides the customer access to applications that would otherwise have been hosted and managed by the customer's own IT staff. Users then access the applications through dedicated intranet connections or over the internet [4],[13]. This allows the customer to focus on their core competencies.

### 2.2. Types of ASPs

ASPs provide services through several models, including Enterprise, Enablers, Pure Play, Vertical, Horizontal, and full service provider (FSP). Enterprise ASPs are typically independent software vendors, such as Salesforce, who offer services directly to the customer. Enablers such as Google support infrastructure used by other ASPs to deliver services. Pure Play providers such as PayPal own the delivered resources and act as the single point of delivery for non-industry specific services such as payment processing. Vertical ASPs like Experion target industry specific applications. Horizontal ASPs provide collaborative or general business applications. FSPs offer end to end solutions.

Table 1. Types of Application Service Providers [14]

| ASP Type | Service Description | Vendors |
|---|---|---|
| **Enterprise** | Independent Software Vendor (ISV) offers their services directly to customer | Sungard, SALESFORCE |
| **ASP Enablers** | Support infrastructure through which ASPs deliver their offerings | ISP or mega resource provider such as AMAZON or GOOGLE |
| **Pure Play** | Own delivered resources & act as single point of responsibility for delivery; often non-industry specific; web deployed applications; e- mail | PAYPAL, Equative |
| **Vertical ASPs** | Targets industry specific applications and processes | Evolved Digital Systems, Experion, Google Healthcare |
| **Horizontal ASPs** | Collaborative applications such as e-mail, or general business oriented applications such as accounting | Hexaware, ConnXion, Apptix |
| **FSP** | Full service providers offering end to end solution | Telecomputing, Inc; Digital Brain; Crimson Logic PTE |

Through the various service models, ASPs provide a multitude of services ranging from enterprise solutions, collaborative applications such as e-mail, business processes such as accounting, to industry specific solutions. VeriSign has even introduced cloud based authentication services [15].

Table 2. ASP Models - Types of ASP Services [16, p. 162]

| ASP Model | Types of Services |
|---|---|
| Collaboration Services | Shared/virtual workspaces, chat/threaded discussion, shared calendar/scheduling, document management, videoconferencing, training /conference registration |
| Electronic Commerce | Auction, catalogs, transaction processing, direct marketing, EDI, logistics, personnel /corporate financial services, surveys, government services |
| Contract Services | Published text, TV and radio programming, games, events, music, community- created content, educational courseware hosting |
| Corporate Systems/ Knowledge Management | ERP, sales force automation, project management, knowledge management, archive indexing, policy-based information access management, claims-processing, e-forms / workflow |
| Interfaces | Portals, search engines, in-box management, personalized information management |
| Networked Smart Products | Copiers, elevators, and other devices that use network connections to send out their own calls for service when required; wireless locator service |
| Infrastructure Outsourcing | Web site hosting, help desk, server/software administration, call centers, directory /addressing /messaging administration, remote management |

## 2.3. History of ASPs

In 2002, Currie & Seltsikas observed that "software [had] evolved from custom coded proprietary applications to pre-packaged, off-the-shelf offerings and [then] to the development of web-centric applications" [16, p. 4]. Currie & Seltsikas also noted that "the convergence of software and IT infrastructure toward a web centric environment" [16, p. 4] facilitated the growth of the ASP concept. ASPs presented a solution to the rapid rate of change in the software market and to the increasingly complex tasks of maintaining, developing, installing and supporting software applications. ASPs provided SMEs a way to reap the benefits of cutting edge technology without large expenditures on IT infrastructure or personnel [4]. ASPs were positioned to provide cost reduction, improved response time, and reduced management difficulties [13], freeing the customer to focus on their core competencies..

Guah & Currie described the early ASP model as "highly volatile, dynamic, and immature" [3, p. 141]. The early ASP market was characterized by under-capitalized start-ups without the resources to survive long term [7]. Few of the early ASPs adopted true web services based platforms, instead clinging to old architecture styles and poor choices of application offerings [4],[7]. Early companies flourished for a few years, but then, in the wake of September 2001, began to flounder. There was too much competition among the early ASPs for too few customers, with little divergence between application offerings and heavy security concerns. Clients found the hosted applications difficult to change or customize. The hosted arrangements were more expensive than anticipated and performance issues with slow broadband networks made most hosted software impractical [10]. But the single greatest contributor to early ASP failure was poor business models and a lack of value creation for the customer [17].

ASPs such as Salesforce and SunGard who repositioned themselves as Software as a Service (SaaS) providers not only survived the early ASP failure, but thrived [10]. Salesforce was founded in 1999 by a group including former Oracle VP and specializes in customer relationship management (CRM) services [18]. SunGard was formed in 1982 from the computer services division of Sun Oil Company and has become the third largest provider of business applications software after Oracle and SAP by acquiring more than 160 companies in order to provide a broad portfolio of best-of-breed technology solutions for enterprise use [19]. The success of both SunGard and Salesforce can be attributed to their focus on core products supporting innate business systems and their use of true internet and web services delivery.

As additional waves of ASPs followed in the SaaS model, industry standards were developed that began to address data security, integration, and interoperability problems using web services. Product offerings began to vary widely and improved broadband speed and reliability made internet delivery a viable option. More flexible pricing options, often by transaction or a monthly access fee, alleviated earlier cost concerns over hosted applications and provided more value to customers.

## 3. CLOUD COMPUTING OVERVIEW

Cloud computing, named for the cloud symbol often used in diagrams and flowcharts to represent the internet, has been made possible by significant innovations in distributed computing and virtualization as well as improved access to higher speed broadband networks [20]. A general consensus agrees that cloud computing includes a dynamic, scalable, pay for use service delivered via the internet that utilizes web based interfaces [7],[20],[21],[22]. According to the National Institute of Standards and Technology, cloud computing is "a model for enabling ubiquitous, convenient, on-demand network access to a shared pool of configurable computing resources … that can be rapidly provisioned and released with minimal management effort or service provider interaction" [22]. Cloud computing could also be considered an evolution of prior ASP models.

Many people compare cloud computing with the utility model for delivery of electricity, as both are metered, dynamic, on demand services. However, unlike electricity, which is measured by the flow of electrons, cloud computing is not measured by data flow. Cloud computing is about service provision, whether computational to collect, manage, or manipulate data; storage, to warehouse or distribute data; or communication, to transport data. A highly consolidated and virtualized data center is not a cloud; automated controls that dynamically assign and continuously monitor physical and virtual resources for efficiency, optimization, and control differentiate a virtualized data center from a cloud.

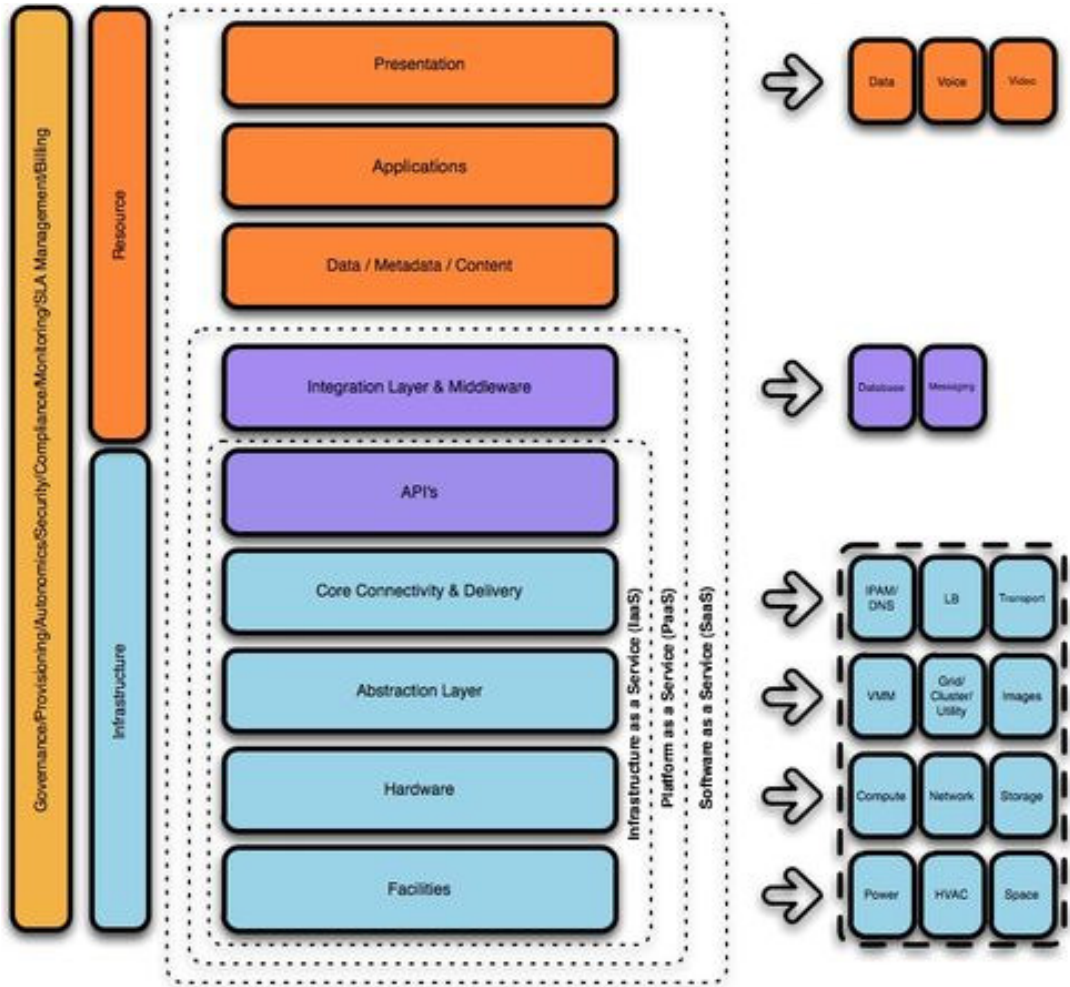

Figure 1. Cloud Ontology & Taxonomy [23].

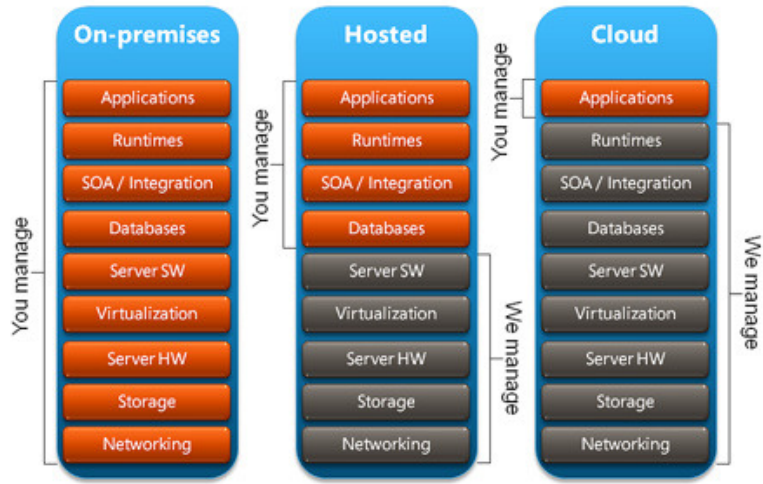

Figure 2. On-Premises, Hosting, or Cloud? [24].

## 3.1. Characteristics of Cloud Computing

The cloud computing model can be described as being composed of five essential characteristics, three service models, and five deployment models [22], [25].

Table 3. Essential Characteristics of Cloud Computing [22, p. 2].

| Characteristic | Description |
|---|---|
| Transparent on demand self service | Consumer unilaterally and dynamically provisions computing capabilities, such as server time and network storage, as needed without requiring human interaction with each service's provider reminiscent of the electricity grid |
| Broad network access | Capabilities are available over the network and accessed through standard mechanisms such as a web browser that are device and location independent. |
| Resource pooling | The provider's computing resources are pooled to serve multiple consumers using a multi-tenant model, with different physical and virtual resources dynamically assigned and reassigned according to consumer demand. There is a sense of location independence in that the customer generally has no control or knowledge over the exact location of the provided resources but may be able to specify location at a higher level of abstraction (e.g., country, state, or datacenter). Examples of resources include storage, processing, memory, network bandwidth, and virtual machines. |
| Rapid elasticity | Capabilities can be rapidly and elastically provisioned, in some cases automatically, to quickly scale out, and rapidly released to quickly scale in. To the consumer, the capabilities available for provisioning often appear to be unlimited and can be purchased in any quantity at any time reminiscent of an electricity grid. |
| Measured Service | Cloud systems automatically control and optimize resource use by leveraging a metering capability appropriate to the type of service. Resource usage can be monitored, controlled, and reported, providing transparency through a utility model. |

## 3.2. Cloud Deployment Models

There are several different types of cloud deployment models, each with their own advantages and preferred use cases. Cloud types include private cloud, virtual private cloud (VPC), public cloud, community cloud, and hybrid cloud. While similar, a highly consolidated and virtualized data center is not a cloud, but is included in the table below to highlight the difference between the models.

While different industries will migrate to the cloud at different rates, most enterprises are actively considering some form of cloud deployment. Many mid to large enterprises are looking at both private and hybrid, while SMEs more heavily consider public clouds due to the low cost of entry. IDC clients were more open to private cloud deployments than public cloud over the next three years [26].

Table 4. Properties of Cloud Deployment Models & Virtualized Data Center [21]

| Property | Virtualized Data Center | Private Cloud | Hybrid cloud | Virtual Cloud | Public cloud |
|---|---|---|---|---|---|
| **Trusted** | High | High | Moderate | Moderate | No |
| **Dynamic** | No | Yes | Yes | Yes | Yes |
| **Controlled** | High | High | Moderate | Moderate | No |
| **Efficient** | No | Yes | Yes | Yes | Yes |
| **Reliable** | Yes | Yes | Yes | Yes | Yes |
| **On Demand** | No | Yes | Yes | Yes | Yes |
| **Secure** | High | High | Moderate | Moderate | No |
| **Flexible** | No | Yes | Yes | Yes | Yes |
| **Cost** | High | High | Moderate | Low | Lowest |

### 3.2.1. Private Cloud

A *private cloud* is operated solely for an individual organization. The cloud can be sited on or off premises and managed by either the organization or a third party. Private clouds have been compared to next generation data centers [22]. Private clouds can provide the automation, provisioning, and deployment orchestration benefits associated with cloud computing while still retaining visibility and control over data [26]. Private clouds are well suited to meet high security, availability, and serviceability requirements related to compliance as well as instances in which data must remain within a certain geographic domain [26].

### 3.2.2. Community Cloud

A *community cloud* is shared by several organizations with shared concerns such as mission, security requirements, policy, and/or compliance considerations [22]. It may be managed by the organizations themselves or by a third party and may exist either on premise or off premise. Using a community cloud allows organizations to benefit from excess capacity within the community rather than having to possibly expand their own data centers to cover occasional excess demand. It can also provide significant environmental and cost savings. Trust, liability, licensing, and other legal and control issues prompt significant concern and present a hurdle to widespread adoption of the model [27]. Implementation of a community cloud model may rely on business interconnect service providers that can act as the central provisioning house for connectivity and ongoing operations [27].

### 3.2.3. Public Cloud

A *public cloud* is made available to the general public or a large industry group by an organization selling cloud services [22]. Most public clouds are multi-tenant, realizing cost savings through shared use of the infrastructure. Separation between users on the public cloud is enacted by use of credentials providing purely logical partitions over an underlying infrastructure shared by many or all users [28]. Public clouds have the lowest entry cost, eliminating capital expenditures on hardware and software, as well as reducing maintenance and operation costs in terms of personnel. Public clouds also have faster implementation times and are the most scalable option. For small to medium enterprises (SME), public cloud use may be the only affordable option. Public clouds are commonly used for workloads such as collaboration, HR, and CRM. Amazon is currently the largest provider of public cloud services with a market share of 80-90% [25].

### 3.2.4. Hybrid Cloud

*Hybrid cloud* infrastructure is a composition of two or more clouds (private, community, or public) that remain unique entities but are connected by standardized or proprietary technology that enables data and application portability [22]. There are multiple scenarios where the utilization of hybrid clouds is more beneficial than a purely private or public implementation. Compliance requirements may require strict controls over data security, but may allow leveraging a SaaS application from a public cloud to process data held in a private cloud; occasional excess demand may require temporary use of capacity from the public cloud, for example seasonal demand for a retailer; or as a disaster recovery option, storing a backup of a server image on a virtual private cloud in case of a disaster for quick conversion to a hot site. Hybrid clouds can offer the flexibility to rapidly change deployments in order to meet changing business needs.

### 3.2.5. Virtual Private Cloud

*A Virtual Private cloud (VPC)* is operated solely for an organization, but located as a subset of a larger cloud infrastructure which may be private, community or public [29]. Most VPCs are logically partitioned, rather than completely physically separated, from the larger cloud. As of April 2011, Amazon also now offers a single tenant, hardware segregated VPC on long term contract for a slight premium to offset the loss of multi-tenant income [30]. Unlike the hybrid model, a virtual private cloud does not interact with other clouds.

According to Amazon VPC services, a VPC can have multiple different compositions. Options include a configuration resembling a traditional networked data center, with isolated subnets, outward facing zones, and even segregated, non-internet connected areas that are directly connected to the enterprise through a VPN [31]. A VPC may also be utilized to store a backup of a corporate data center on a few virtual instances for disaster recovery purposes at extremely low cost [31]. Then, in the case of a disaster to the data center, a full, scalability implementation can quickly be launched until the data center is restored. When the data center is restored, the revised image can be ported back to the data center, releasing the leased capacity. This provides rapid hot site capability at a fraction of the cost of a traditional infrastructure project.

### 3.3. Cloud Service Models

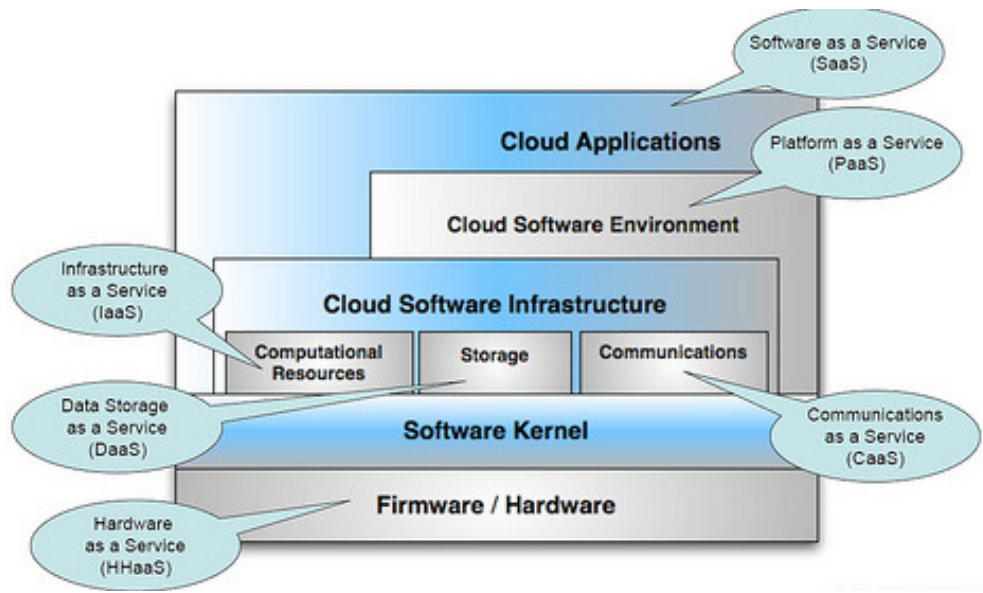

Figure 3. Cloud Service Layers [30, p. 13].

Most descriptions of cloud services identify three primary models: Infrastructure as a Service (IaaS), platform as a service (PaaS), and software as a service, very similar to the previously identified ASP service models. van Eijk and Yousef take the taxonomy of the cloud further, and more accurately describe the variety of implementations available from the cloud [25],[32]. At the most basic level, hardware as a service (Haas) offers access to hardware, firmware, and operating system. At the next level, infrastructure services are available. Here, Yousef and van Eijk are more concise, expanding the model to include Data Storage as a Service (DaaS) and Communications as a Service (CaaS) [25],[32]. On the next level, platform as a service (PaaS), allows customers to create their own individuals cloud applications. And at the highest level, software as a service (SaaS) provides seamless application delivery. These separated offerings are particularly important within hybrid environments, where compliance requirements may require organizations to utilize private clouds for data storage but permit program execution through SaaS, PaaS, or IaaS on a public cloud. In some cases, however, components may cross definitional lines. For example, a CaaS product in some implementations may be more of a software service than an infrastructure service, or may include features of both.

In addition to the primary IaaS, CaaS, DaaS, PaaS, and SaaS vendors, numerous middleware vendors have emerged to provide applications and software that make managing and operating cloud services easier, promote self service, or provide key architectural pieces for developers to launch their own services. Middleware vendor Apprenda offers an application server built specifically to solve architecture, business, and operational complexities of delivering SaaS in the cloud, while AppDynamics offers management services for application architectures that span the cloud or data center to monitor, troubleshoot, diagnose, and scale production across multiple platforms, and CloudOptix offers virtualization software to launch multivendor public and private clouds and then unify them into a custom hybrid to maximize cost savings [33]. Other vendor offerings include cloud gateways, dynamic cloud usage during excess resource demand periods (referred to as cloud bursting), cloud automation solutions, and performance level testing.

Table 5. Cloud Service Categories

| | | | | |
|---|---|---|---|---|
| Vendors | Amazon EC2<br>AT&T Synaptic<br>CA<br>Cloudscaling<br>GoGrid<br>HP<br>Rackspace<br>Verizon | Amazon<br>Flexiant<br>FORCE.COM<br>Google Apps<br>gCloud3<br>MS Windows Azure<br>OpenStack<br>RightScale | Gordano<br>Hotmail<br>Intacct<br>MaxxCloud<br>MS Office 365<br>NetSuite<br>Now2Office<br>SalesForce | Akamai<br>Amazon Cloudfront<br>Axcient<br>Carbonite<br>Cleversafe<br>Doyenz<br>Intronics<br>Rackspace Cloudfiles |
| Control | Customer retains control over operating systems, storage, deployed applications, and possibly limited control of select networking components | Customer retains control of deployed actions & possibly application hosting environment configuration | Customer does not manage or control underlying infrastructure | Customer does not manage or control underlying infrastructure |
| Service | Virtual server instances with unique IP addresses and blocks of storage on demand activated by provider's application program interface (API) to start, stop, access, and configure their virtual servers and storage.<br>Capacity is dynamically allocated or released as required with client paying for only as much capacity as needed. | Hosted software and product development tools that developers utilize to create applications on the provider's platform over the Internet using APIs, website portals or gateway software.<br>No current standards for interoperability or data portability; some providers prohibit removal of software created by their customers from the provider's platform | Software product interacts with the user through a front-end portal.<br>Services can be anything from web-based email to inventory control and database processing<br>End user is free to use the service from anywhere as the service is device and location independent. | This model includes data storage, backup services, business continuity, and static content distribution networks (SCDN).<br>SCDN is a behind the scenes service to distribute content faster and more efficiently through the use of a network of geographically dispersed servers. Netflix is an example of customers who utilize SCDN. |
| Main cost unit | Per virtual server | Per unit API call or chunk of code | Per user account per month | Per storage unit (single image, movie, Gb, etc) |

Some may ask is the cloud real, or is it just a convenient advertising hype? Indeed, the cloud is real in the sense that companies who provide these types of services exist and their offerings can be measured in terms of growth and performance. Analysis of these metrics is an important step when selecting a cloud infrastructure vendor. The main cost unit for vendors such as Amazon's Elastic Compute Cloud (EC2) is the virtual server [25]. van Eijk performed tests on cloud uptime, responsiveness, and proximity. In calculating the responsiveness of various services compared to a standard New York based website, the average number of milliseconds per roundtrip among 35 worldwide monitoring stations was examined. According to van Eijk, Akamai "is everywhere" and Google has 36 worldwide data centers comprising more than 1 million servers [25]. van Eijk observed an overall improvement in metrics in almost every category between April and September of 2009 [25], supporting his conclusion that cloud services are continuing to expand and improve over time.

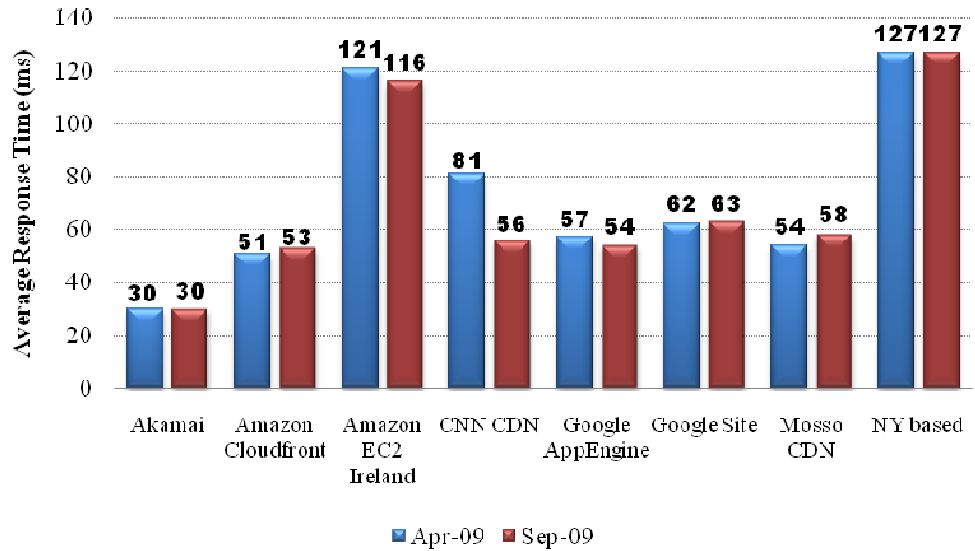

Figure 4. Cloud Proximity by Average Response Time [23, p. 11]

Another consider when selecting a cloud provider is that each provider has strengths and weaknesses. Amazon lacks migration compatibility and gross size availability for very large deployments between their limited number of physical sites, whereas Google, while still in beta, has a large number of worldwide sites that are consistent for migration with greater capacity with duplication over distance.

## 4. BENEFITS & RISKS OF ASP USAGE

ASP usage has both benefits and risks that must be carefully considered. One of the most significant benefits of ASP usage is reduced cost. A consistent monthly lease fee eliminates infrastructure development, implementation, maintenance and support costs related to maintaining an in house computing infrastructure as well as costs related to maintain the latest technology. Other benefits include obtaining expertise in a particular field without the expense to retain skilled information professionals, lack of risk from undertaking custom development, flexibility of short term contracts without long term commitments, scalability, interoperability, speed, and portability. The most serious risks with ASP usage include dependence on an ISP or provider out of direct control of the enterprise, vulnerability of confidential data and processes, and the loss of service if the provider were to cease operations. Other risks include a lack of specialization and customization from generic offerings, internet bandwidth usage and availability, and integration with legacy systems.

Table 6. Benefits & Risks of ASP usage

| | | |
|---|---|---|
| **Cost** | Monthly lease fee or per transaction fee eliminates infrastructure, development, maintenance, and support costs providing more consistent pricing. Instead of capital expense, now operating expense | |
| **Customization** | Some ASPs provide customization and integration options (Google APPS, Salesforce), but are limited | Most ASP offerings are very generic, with little opportunity for custom development. |
| **Flexibility** | Short term contracts can allow termination at short notice; scalability; ease of entry | |
| **Integration** | | Difficult to integrate legacy systems, especially for large firms |
| **Interoperability** | Seamless connectivity and integration among diverse business partnerships; hardware independent, and in some cases even available from mobile devices | Lack of standards, so applications may not be interoperable between vendors. |
| **Portability** | Hosted application(s) available from any workstation with internet access; may even be available from mobile devices | Some vendors do not allow applications to be removed from their platform |
| **Reliability** | High capacity and redundancy by dedicated provider may have improved uptime stats for 24/7 operations, especially for SME compared to in-house ability | Dependent on ISP or infrastructure provider out of ASP or customer control. Risk of provider going out of business. |
| **Security** | Experienced vendor may provide more security, and be able to research compliance regulations, more effectively than SMEs | Vulnerability of confidential data and processes within a 1-many environment; underlying security concerns of web environment from hacking, governance control |
| **Speed** | Adoption / implementation time faster than vendor or in-house development | Internet bandwidth is inconsistent. Some SMEs, particularly in rural areas, do not have access to fiber optic cabling or are reliant on older technology [34]. |
| **Technology** | Able to adapt quickly to changing technology without large capital or manpower investment; access to highly skilled IT professionals | |

## 4.1. ASP Target Market

The ASP model is most successful targeting multiple customers for a single generic application, or suite of applications, requiring little customization. The more customization required, the more difficult it is for the ASP to create cost savings from a volume business.

SMEs do not have the financial resources of larger enterprises to invest in IT infrastructure or personnel, and are a prime customer base for ASP usage. SMEs also have fewer legacy and integration issues than larger enterprises who tend toward large legacy systems with tremendous amounts of data requiring conversion or adaptation [10].

### 4.1.1. ASP Adoption Example – United Capital Financial Advisers

Crosman reported the example of a moderately sized investment firm, United Capital Financial Advisers, based in California with 15 offices spread throughout the country [10]. United grew by acquiring other smaller firms and wanted a single, centralized Customer Relationship Management (CRM) system that would be easy for new acquisitions to adopt. Each new acquisition had their own CRM system from a variety of sources, but none met United's requirements, including an easy conversion path that was scalable in the face of new acquisitions. Instead of becoming a software developer, United sought a well-built, reliable solution that was easy to configure, which they found in Salesforce. What the United staff could not do to configure the software, Salesforce did very responsively, ultimately allowing United to integrate several other business systems into the Salesforce platform. With Salesforce's month to month pricing, United did not feel locked into a single vendor; Salesforce was constantly forced to prove themselves and United could leave at any time [10].

## 5. DISTRIBUTED COMPUTING & VIRTUALIZATION

Utility computing was first suggested by Stanford computer scientist Dr. James McCarthy during a 1961 speech celebrating MIT's centennial. Dr. McCarthy proposed a future where computing time and even specific applications could be sold in a pay per use model like water or electricity [35].

Distributed computing is one of the paradigms used to implement the idea of utility computing. Dupre compared it to "*cluster computing (*emphasis original) which views a group of linked computers as a single virtual computer for high-performance computing, or *grid computing,* (emphasis original) where the linked computers tend to be geographically distributed to solve a common problem" [33, p. 1].

Virtualization is used to partition large computing resources into multiple smaller resources on demand, providing scalability and transparency. The user can then utilize only the amount of computing truly needed, releasing unused resources to other users. Virtualization also increases reliability by encapsulating each instance into separate environments which are not affected by the operation of other instances. If a virtual instance fails, only that instance needs to be restarted rather than the entire machine. Virtualization and utility computing are prime components of cloud computing.

## 6. BENEFITS & RISKS OF CLOUD COMPUTING

Enterprises continue to expand beyond traditional IT boundaries, with 75% of employees accessing computing services outside the enterprise perimeter at least part time [36]. As that use continues to climb, cloud computing is becoming increasingly attractive. The most obvious benefits of cloud computing are scalability, agility, device and location independence, transparency, portability, efficiency, cost savings, and innovation. Reliability, with excellent fall over capabilities and fast portability is a potential benefit, but cloud providers have also experienced total outages.

Scalability is a primary benefit of Cloud Computing. Companies are able to pay for only the amount of computing resources they actually consume at any given time with the ability to increase or decrease those resources at any time. This type of flexibility can be especially relevant to SMEs or startups with varying amounts of usage or growth potential to avoid either over or under capacity planning. One example is the small animation company Animoto who experienced a tremendous, temporary surge in demand from a Facebook review that would have crippled a traditional in-house implementation [37].

Agility is a significant motivator in the adoption of cloud computing by larger enterprises. Unlike traditional in house data centers and development, cloud services can be brought on line in as little as a few minutes, instead of the weeks or months required in traditional deployments [38]. Cloud computing can rapidly provision computing resources to support real time business needs, supporting innovation and allowing enterprises to more quickly respond to changing business conditions or opportunities.

Transparency, device and location independence, and portability benefits of cloud computing support the changing nature of the enterprise. With more employees operating outside of the traditional boundaries of the enterprise, cloud computing allows those services to follow the employee, often right to customer themselves. These benefits can also support increased agility for the enterprise to respond to changing conditions and to innovate to provide their customers better, more effective services and products tailored to their needs.

Infrastructure companies such as Boomi, with their Atomsphere offering, are lowering the integration burden for cloud adoption. Atomsphere allows organization to connect cloud and on premises applications without software or appliances [39].

## 6.1. Cost Savings with Cloud Computing

In terms of direct infrastructure and IT support costs, cloud computing is significantly less expensive for an enterprise than to build out and operate an internal data center [40]. One might compare cloud computing to a utility model – in most cases, it is significantly less expensive to acquire electricity from a utility company than to build and operate a dedicated power generation plant. In an expanding market, usage of the cloud can mean immediate access to increased resources, rather than months, or years, required to expand an in house data center. The US General Services Administration (GSA) moved USA.gov to the cloud in order to meet increased usage requirements. Forecasted growth requirements on the cloud require a one day upgrade at a cost of $806,000 annually, compared to six months and $2.47 million annually using traditional IT procurement, realizing a savings of 67% [41]. HP estimates it can save enterprises 56% over traditional data center costs through implementation of a private cloud [33]. NASA CIO Linda Cureton notes that cloud computing provides opportunities to reduce fixed costs as well as an alternative to the high entry costs of in-house development for government agencies while Casey Coleman, CIO of GSA, is more bullish, saying that "Cloud computing [has] come of age" [41].

Centralization can also lower infrastructure costs even for private clouds by locating cloud data centers in areas with lower real-estate or electricity costs. Centralization through leased services, with either a public or private cloud provider, can even achieve no physical footprint at all. In some locations worldwide, such as the Netherlands, creation of new physical datacenters is under moratorium because of power consumption requirements [25]. In essence, clouds could be said to be more energy efficient as many organizations can share processing capability on a shared basis, rather than wasting unused instances and resources in order to prepare for peak usages.

Utilizing cloud based services can save even large companies on expensive infrastructure investments. VeriSign, for example, offers a cloud based strong authentication system where the infrastructure is provided by VeriSign, but credentials administration can be performed in house. It is estimated that their cloud based services will provide total cost of ownership savings of up to 40% compared to an in-house implementation, and reduce ongoing IT support costs by 90% [36]. And for SMEs, cloud based services remove the large entry barrier for utilizing more advanced technology by eliminating high startup investment, and can results in even more significant savings for an organization just implementing strong authentication.

SCDN services through companies such as Akamai or Rackspace can improve the responsiveness of an e-commerce site, increasing profitability and order throughput. Akamai estimates a large savings in infrastructure costs by offloading 75% of server acquisitions, software acquisition, and operations to their distributed network [42]. ROI is observed from both infrastructure savings and improved conversion rates. Savings vary by each enterprise's unique requirements, so Akamai provides an ROI calculator [42]. For example, a SME anticipating 5 web servers and 2 application servers with 10,000 hits per month, an average transaction amount of $50, 3% conversion rate and 40% GMP, estimated ROI savings are in excess of $425,000 per year [42]. For a larger organization, the savings can amount to millions of dollars.

## 6.2. **Risks with Cloud Computing**

Serious security concerns still remain about cloud computing. Major risks can be grouped into six categories, none of which are unique to the cloud, but are rather amplified by cloud architecture, include [43]:

1. Abuse and nefarious use
2. Malicious insider risks
3. Account, service, and traffic hijacking [authentication and access tracking]
4. Insecure application programming interfaces (APIs)
5. Shared technology vulnerabilities
6. Data loss and leakage

In 2009, T-Mobile Sidekick users irrecoverably lost data when a storage array was upgraded without first being backed up [44]. Simple human error, but the scale of cloud computing compounded it, as well as complacency that the cloud was the backup and did not, itself, require redundancy. Lawton & Whitney report that a serious security breach occurred in December 2009 at Amazon with a botnet incursion intending to capture banking data from customers [45],[46]. And a data breach of Epsilon in March 2011 that exposed customer's e-mail addresses and other personal information from over 150 major banks and retailers [47], was eventually traced back to an incursion into RSA. Other major incursions are expected to occur, and it is only a matter of time before hackers begin forming their own cloud infrastructures as they reinvest cybercrime profits into new technologies [45]. Security standards being developed by the Cloud Alliance and the European Network and Information Security Agency will help address some of these vulnerabilities, but it will be some time before companies feel secure enough to trust the cloud with their most sensitive data [1],[45].

Other concerns about cloud computing include the ability to audit and monitor the service. The Stored Communications Act (SCA) limits the enterprise's ability to manage the information resource in situations such as a password protected e-mail account. While the service was provided by the enterprise, they retained the right to monitor or audit during the normal course of business. When a third party provides the service, the enterprise no longer has that right or control. Where control is a major issue, many organizations are looking to private or hybrid cloud adoption which do not generate as much cost savings as a public cloud, but offer the enterprise more direct control and are more likely to meet compliance requirements.

While security and control are major concerns regarding cloud adoption, network capacity, availability, performance, viability, interoperability, consistency and standards are also concerns for enterprises just as they were during the early ASP adoption. Network capacity and composition is a major concern towards cloud adoption, just as it was to ASP adoption. Increased usage of the cloud through internet connectivity places a higher burden on the network infrastructure of the enterprise, and with an increasing proportion of employees working remotely, both the composition and boundaries of the network may require optimization [26]. Increased capacity needs may require increased infrastructure investment and modification of network architecture in order to maintain acceptable performance levels, reducing the initial perceived cost savings of cloud implementation. Networks themselves may also require adjustment in order to become more responsive and scalable.

Viability of the cloud provider must be considered, specifically their ability to provider to ensure continuation of services, which may be lost due to financial or legal difficulties. Amazon's 80-90% market share in public cloud services [25] is being eyed hungrily by many new entrants to the field, including giants AT&T, IBM, and Verizon, as well as vendors who only play in the private cloud market, so competition should keep pricing down, but also run the risks that destroyed the ASP market – over competition.

Cloud development is still in its early stages, and while some APIs from vendors are well known and available under Creative Common's licensing, and there are early movements towards creation of standards, most development environments are not interoperable between vendors. So while applications may be portable from device to device, they are not likely to be portable between cloud service vendors. RightScale is attempting to prevent lock-in by providing visibility into all levels of deployment and permitting users to choose their deployment language, environment, stack, data-store and cloud for portability [39].

## 7. Cloud Computing Target Market

Cloud computing has been called "supercomputing for the masses" [11]. SMEs, particularly those with multiple physical locations, are a prime market for cloud computing, as they were for ASPs. Certain industries are more receptive to cloud adoption, particularly where immediate data exchange is critical. The financial services industry is a heavy adopter, second only to the technology industry [10]. New healthcare regulations requiring the portability of patient data are pushing early adoption of cloud computing into the healthcare industry. With their low margins, many healthcare providers cannot afford large investments in IT infrastructure [6]. And government, with its increasing dependence on commercial off the shelf software and need to slash costs, is leading as a heavy cloud adopter. GSA is even marketing cloud services to other government departments [48].

Many companies are considering adoption of cloud computing; in 2009, Johna Johnson, president of Nemertes Research, stated that while only 2% of her clients had currently adopted cloud computing, 20% were considering it, and the majority of IDC clients are considering cloud adoption [49], [26]. Most telling, perhaps, is a survey by Tier 1 research indicating that 60% of SME's are planning to implement either IaaS or SaaS offerings within the next 12 months [50].

Large enterprises, however, have been slow cloud adopters. Security concerns may not be the biggest challenge to the adoption of cloud computing within large enterprises; the culture of the large enterprise itself, combined with enormous investments in traditional architecture, may be the most difficult hurdle to overcome [51]. While enterprise adoption of cloud services has been slow, estimates place enterprise data center migration to the cloud to reach 35% by 2015 [50].

## 8. CONCLUSION

It could be argued that cloud computing is nothing new. Decades earlier mainframes provided portability through remote access terminals and ASPs have provided application hosting for more than a decade. Rather, cloud computing is a convergence of multiple IT strategies, technological advances, and the willingness to embrace open standards in a way that has come together to provide true utility computing as envisioned by Dr. James McCarthy nearly half a century ago.

Cloud computing is exploding at a remarkable rate. Within the next decade, experts such as Diffie speculate that isolated computing will be at an end [1]. Enterprises, large and small, will become dependent on a conglomeration of companies providing computational services, likely in the form of cloud computing, faster and cheaper than the enterprise could themselves, much as companies have already become dependent on search engines such as Google [1].

Many ASPs, such as Salesforce and SunGard, who adopted the SaaS concept and moved their infrastructures to the web services standard have already made the transition to the cloud. Other ASPs still mired in outdated architectural and deliverable formats will have a much more difficult transition to cloud computing, if at all, and may pass into history like those companies in previous waves who were not adaptable to the changing marketplace. Overall, with the divergence of offerings, increasing network capacity, and better business plans, cloud vendors are overcoming the pitfalls that doomed many ASP vendors in the past. Cloud computing is maturing into a new paradigm. Much like client server architecture denoted a move from the dominance of mainframes, cloud computing is now moving beyond the enterprise data center.

## AUTHORS

**Kathleen Jungck** is a graduate student in Information Security and Assurance at Capella University.   She has worked in many facets of the Information Technology industry for employers including the University of Washington and Boeing Interiors.

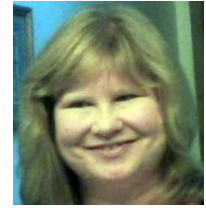

**Syed (Shawon) M. Rahman** is an assistant professor in the Department of Computer Science and Engineering at the University of Hawaii-Hilo and an adjunct faculty of information Technology, information assurance and security at the Capella University. Dr. Rahman's research interests include software engineering education, data visualization, information assurance and security, web accessibility, and software testing and quality assurance. He has published more than 50 peer-reviewed papers. He is a member of many professional organizations including ACM, ASEE, ASQ, IEEE, and UPE.

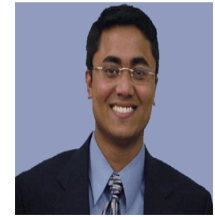